\documentstyle[osa,aps,prl,epsfig]{revtex}
\begin{document}
\twocolumn[\hsize\textwidth\columnwidth\hsize\csname
@twocolumnfalse\endcsname
\title{%$^{\ast\ast\ast}$
Monoclinic and Correlated Metal Phase in VO$_2$ as Evidence of the
Mott Transition: Coherent Phonon Analysis}
\author{Hyun-Tak Kim$^{1\ast}$, Yong-Wook Lee$^1$, Bong-Jun Kim$^1$, Byung-Gyu Chae$^1$, Sun Jin Yun$^1$, Kwang-Yong Kang$^1$, Kang-Jeon Han$^2$, Ki-Ju Yee$^2$, and Yong-Sik Lim$^3$}
\address{$^1$IT Convergence and Components Research Lab., ETRI, Daejeon 305-350, Republic of
Korea\\$^2$Department of Physics, Chungnam National University,
Daejeon 305-764, Republic of Korea\\$^3$Department of Applied
Physics, Konkuk University, Chungju, Chungbuk 380-701, Republic of
Korea}
\date{June 22, 2006}
\maketitle{}
%\newpage
\begin{abstract}
In femtosecond pump-probe measurements, the appearance of coherent
phonon oscillations at 4.5 THz and 6.0 THz indicating the rutile
metal phase of VO$_2$ does not occur simultaneously with the
first-order metal-insulator transition (MIT) near 68$^{\circ}$C.
The monoclinic and correlated metal(MCM) phase between the MIT and
the structural phase transition (SPT) is generated by a
photo-assisted hole excitation which is evidence of the Mott
transition. The SPT between the MCM phase and the rutile metal
phase occurs due to subsequent Joule heating. The MCM phase can be
regarded as an intermediate non-equilibrium state.

\\
\end{abstract}
]
%\newpage
%\narrowtext
%\section{INTRODUCTION}
\newpage
New physical phenomena such as high-$T_c$ superconductivity,
colossal magnetoresistance and dilute magnetism occur when
strongly correlated materials are doped with hole charges. These
phenomena may be caused by a strongly correlated Mott first-order
metal-insulator transition (MIT) with a change in on-site Coulomb
interaction without the accompanying structural phase transition
(SPT) \cite{Mott}. The Mott MIT is still under intense debate,
even though many scientists have tried to clarify the MIT
mechanism \cite{Imada}. We expect that studies on the MIT may
provide a decisive clue in understanding these new phenomena.

VO$_2$ (paramagnetic $V^{4+}$, 3$d^1$) had monoclinic M$_1$,
transient triclinic T, monoclinic M$_2$, and rutile R phases
\cite{Pouget}. The monoclinic phase has two electronic structures,
which are one-half of the V chains of the R phase being paired
without twisting, while the other half twist but do not pair
\cite{Pouget,Rice,Tomakuni}. The M$_2$ phase, consisting of
equally spaced V chains, was defined as a monoclinic Mott-Hubbard
insulator phase and the M$_1$ phase, insulating monoclinic phase,
may be a superposition of two M$_2$-type lattice distortions
\cite{Pouget,Rice}. The first-order MIT was considered to be due
to a change in atom position \cite{Pouget,Rice}. More recently,
Laad $et~al.$ calculated, using the local-density approximation +
dynamical mean field multiorbital iterated-perturbation theory
scheme, that the MIT from R to M$_1$ phases was accompanied by a
large spectral weight transfer due to changes in the orbital
occupations \cite{Laad}. This supported the Mott-Hubbard picture
of the MIT in VO$_2$, where the Peierls instability arises
subsequent to the MIT \cite{Kim-2,Kim-3}.

For SPT in VO$_2$, in contrast, it has also been proposed that the
MIT near 68$^{\circ}$C is the Peierls transition caused by
electron-phonon interaction. For example, it has been argued that
VO$_2$ is an ordinary band (or Peierls like) insulator on the
basis of the $d_{II}$-bonding combination of the 3$d^1$ electron,
resulting in a Peierls-like band gap \cite{Goodenough}. The same
conclusion was also reached by band structure calculations based
on the local density approximation \cite{Wentzcovitch}, an
orbital-assisted MIT model \cite{Haverkort}, a structure-driven
MIT \cite{Cavalleri-1}, and experimental measurements of a
structural bottleneck with a response time of 80 fs
\cite{Cavalleri-2}. This has been also supported by other models
including electron-electron interaction and electron-phonon
interaction \cite{Biermann,Okazaki}.

These controversial argue on the electronic structure of VO$_2$ is
largely due to preconception that the MIT and the SPT occur
simultaneously even though there may only be a causal relation
between the SPT and the MIT.

We developed theoretically the hole-driven MIT theory (an
extension of the Brinkman-Rice picture \cite{Brinkman}) with a
divergence for an inhomogeneous system \cite{Kim-1,Kim-4}, and
have reported the MIT with an abrupt first-order jump in
current-voltage measurements below 68$^{\circ}$C, the SPT
temperature \cite{Kim-2,Kim-3,Chae,Kim-4}. We have demonstrated
that the Mott MIT occurs when the valence band is doped with hole
charges of a very low critical density \cite{Kim-2,Kim-1,Kim-4}.

In this letter, we report time-resolved pump-probe measurements
for VO$_2$ and demonstrate the causality between the MIT and the
SPT by analyzing coherent phonon oscillations that reveal
different active modes across the critical temperature. We also
simultaneously measure the temperature dependence of the
resistance and crystalline structure to confirm the optical
results. To our knowledge, this letter is the first to report the
simultaneous analysis of MIT and SPT of VO$_2$. We newly define a
$\bf m$onoclinic and $\bf c$orrelated $\bf m$etal(MCM) phase
between the MIT and the SPT, which is different to Pouget's M$_2$
definition \cite{Pouget,Rice}, on the grounds that the MIT occurs
without an intermediate step. Photo-assisted temperature
excitation measurements that were used to induce a new MCM phase
as evidence of the Mott transition are also presented. The origin
of the MCM phase is discussed on the basis of the hole-driven MIT
theory \cite{Brinkman,Kim-1,Kim-4}. The use of this theory is
valid because the increase of conductivity near 68$^{\circ}$ was
due to inhomogeneity \cite{Choi} and inhomogeneity in VO$_2$ films
was observed \cite{Kim-2,Kim-3,Chae,Kim-4}.

High quality VO$_2$ films were deposited on both-side polished
Al$_2$O$_3$(1010) substrates by the sol-gel method \cite{Chae-2}.
The thickness of the films is approximately 100 nm. The
crystalline structure of the films was measured by x-ray
diffraction (XRD). For coherent phonon measurements, time-resolved
transmissive pump-probe experiments were performed using a
Ti:sapphire laser which generated 20 fs pulses with a 92 MHz
repetition rate centered on a wavelength of 780 nm. The diameter
of the focused laser beam was about 30${\mu}$m.

\begin{figure}
\vspace{-0.6cm}
\centerline{\epsfysize=13cm\epsfxsize=9cm\epsfbox{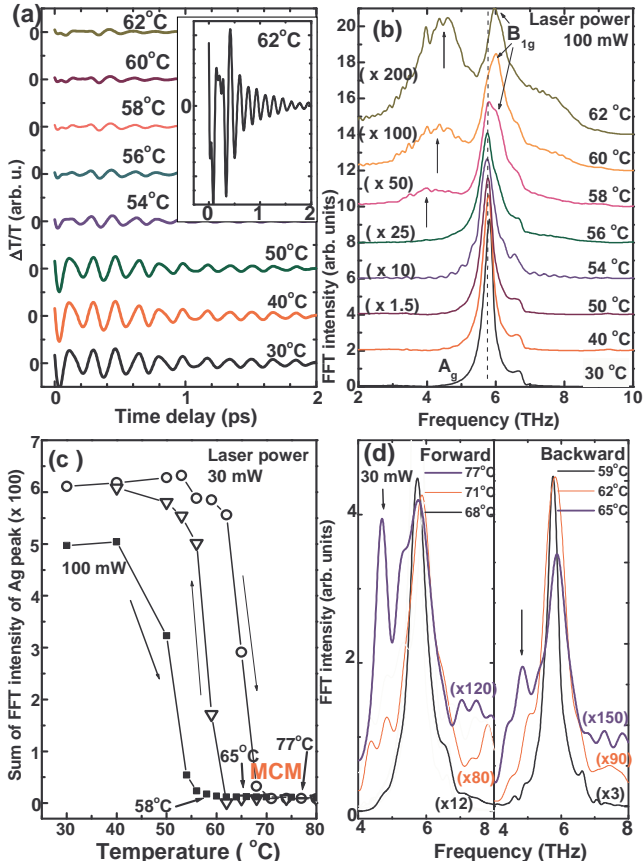}}
\vspace{-0.2cm} \caption{(color online) (a) The temperature
dependence of coherent phonon oscillations measured in a time
domain at a pump power of 100mW. (b) The temperature dependence of
the fast Fourier transformed (FFT) spectra taken from the coherent
phonon oscillations in Fig. 1(a). The peaks at 4.5 and 6 THz
appear simultaneously. (c) The temperature dependence of the sum
of the monoclinic A$_g$ peak spectral intensities measured at pump
powers of 30 and 100mW. (d) The temperature dependence of FFT
spectra of coherent phonon oscillations measured in the forward
and backward directions at 30mW pump power near $T_{SPT}$.}
\end{figure}

The coherent phonon measurement has many advantages over
conventional continuous wave spectroscopy, such as the
amplification effect of coherent phonons and a very low level
background in the low wavenumber range \cite{Yee}. Figure 1(a)
shows the temperature dependence of coherent phonon oscillations
measured at a pump laser power of 100mW. The enlarged oscillation
trace measured at 62$^{\circ}$C is shown in the inset. The
oscillation traces below 54$^{\circ}$C are clear, while for
temperatures of 54$^{\circ}$C and above the oscillation amplitude
is weakened since the MIT has already occurred.

Figure 1(b) shows the temperature dependence of coherent phonon
peaks obtained by taking a fast Fourier transform (FFT) of the
time-domain oscillations in Fig. 1(a). The intense monoclinic
A$_g$ peak near 5.8 THz decreases in intensity as the temperature
increases. The A$_g$ peak finally disappears at 58$^{\circ}$C. A
new broad peak near 4.5 THz (150 cm$^{-1}$) and a sharp peak at
6.0 THz (202 cm$^{-1}$) appear at 58$^{\circ}$C and over, as
denoted by the arrows in Fig. 1(b). The intense peak at 6.0 THz is
identified as the B$_{1g}$ (208 cm$^{-1}$) Raman active mode of
the R phase \cite{Srivastava}, but the broad peak near 4.5 THz
(150 cm$^{-1}$) is not assignable, because this mode is excluded
from allowable Raman active modes of the rutile structure
\cite{Srivastava,Schilbe}. We suggest that this peak belongs to an
active mode of the R phase because it appears at the same time as
the B$_{1g}$ mode. Moreover, it is obvious that the large decrease
of the A$_g$ peak is attributed not to the SPT but rather to the
MIT approximately between 50 and 58$^{\circ}$C.

Figure 1(c) shows the temperature dependence of the sum of the
A$_g$ peak spectral intensity centered at 5.8 THz from FFT spectra
in Fig. 1(b). These measurements were performed with pump powers
of 30 mW and 100 mW, respectively. The temperature dependence of
the spectral intensity shows a similar trend irrespective of pump
power. The hysteresis curve at 30 mW is denoted by circles
(heating: forward) and triangles (cooling: backward). The circle
curve with the intensity drop near 68$^{\circ}$C is similar to the
resistance curve of VO$_2$. This result indicates that the heating
effect due to a focused laser beam at 30mW is negligible.
Conversely, the transition temperature observed for a pump power
of 100 mW (black filled squares) is $\sim$12$^{\circ}$C lower than
that observed for the 30 mW case (Fig. 1 (c)). This is due to a
local heating effect by laser spot \cite{Laser}. Thus, the true
temperature of 58$^{\circ}$C shown in Fig. 1(b) is likely to be
$T_{SPT}$=70$^{\circ}$C.

Figure 1(d) shows the coherent phonon spectra obtained for forward
(heating) and backward directions at 30 mW. Remarkably, the
coherent phonon peak at 4.5 THz appears at 77$^{\circ}$C (left
side in Fig. 1(d)). The phonon peaks near 71$^{\circ}$C indicate
an intermediate state for the SPT. The phonon peaks at
74$^{\circ}$C (not displayed in the figure) showed the same
behavior as those at 71$^{\circ}$C. The SPT temperature is
regarded as 77$^{\circ}$C. Thus, the MCM phase is in a temperature
range from 62 to 77$^{\circ}$C (Fig. 1(c) and 1(d)). Although a
temperature increase due to the low pump power is insignificant,
the laser can still excite holes in the film \cite{Rini}. The
photo-induced holes cause the transition to the MCM phase, which
will be discussed in a following section. In the backward
direction of the hysteresis curve (denoted by triangles in Fig.
1(c) and right side in Fig. 1(d)), the peak at 4.5 THz is
recovered at 65$^{\circ}$C and the intermediate structure also
appears at 59$^{\circ}$C. The corresponding $T_{SPT}$ in the
cooling cycle is then regarded as being 65$^{\circ}$C. Such a
decreased $T_{SPT}$ (77$\rightarrow$65$^{\circ}$C) is due to a
residual heating effect from the high temperature increase up to
80$^{\circ}$C. For the cooling process, the MCM phase is likely
present in a temperature range between 56 and 62$^{\circ}$C.

\begin{figure}
\vspace{-0.1cm}
\centerline{\epsfysize=11cm\epsfxsize=8.4cm\epsfbox{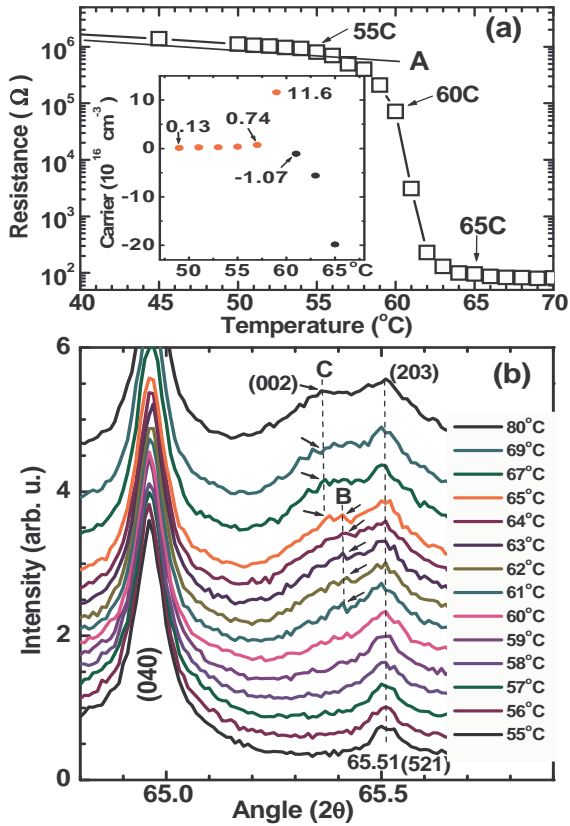}}
\vspace{0.2cm} \caption{(color online) (a) The temperature
dependence of the resistance simultaneously measured by x-ray
diffraction. The inset cited from reference 6 shows the
temperature dependence of carriers near the MIT. Red bullets
indicate holes. (b) The temperature dependence of the XRD data.}
\end{figure}

In order to confirm the above optical results, we simultaneously
measured the temperature dependence of the resistance and XRD
pattern for the VO$_2$ (521)/Al$_2$O$_3$(1010) films. The
temperature range shown in Fig. 2(a) corresponds to the
temperatures of the XRD measurement in Fig. 2(b) and is performed
at 1$^{\circ}$C intervals. At 56$^{\circ}$C, the resistance begins
to deviate from the linear fit denoted by line A. $T_{MIT}$ can be
regarded as 56$^{\circ}$C. The (521) plane (2${\theta}$
=65.42$^{\circ}$) in Fig. 2(b) has a single peak at 55$^{\circ}$C.
A shoulder on the (521) plane peak appears at 61$^{\circ}$C, as
indicated by dot-line B in Fig. 2 (b). The peaks indicated by line
B denote an intermediate structure and not the R metal phase. The
peak related to the (002) plane appears at 65$^{\circ}$C and
represents the R metal phase, as indicated by dot-line C in Fig. 2
(b). It is likely that the SPT begins at 65$^{\circ}$C and is
continuous with temperature. This is the same result as in
previous work \cite{Leroux}. As shown in Fig. 2(a), the resistance
at 65$^{\circ}$C is as small as about 100${\Omega}$ above
65$^{\circ}$C. This is four orders of magnitude lower than the
resistance of the insulating phase. These results clearly show
that $T_{MIT}$ is different to $T_{SPT}$ and that the MCM phase
exists between 56 and 65$^{\circ}$C as an intermediate state.

Furthermore, the inset in Fig. 2(a) shows a Hall measurement in
which a change of carriers from holes (red bullets) to electrons
(black bullet) is observed near 60$^{\circ}$C. This indicates that
the type of carrier in the metal phase is an electron. This shows
that the MIT has already occurred at 60$^{\circ}$C lower than
$T_{SPT}$, and that hole carriers drive the first-order MIT
\cite{Kim-2}. It should be noted that the hole-driven MIT is quite
different from the well-known original Mott transition idea where
a first-order MIT from a metal to an insulator occurs when
electron carriers in the metal are reduced to a critical density.
This is caused by a reduction of the screened long range Coulomb
potential energy. However, there is no change of carrier type in
Mott's idea.

A femtosecond X-ray study for the SPT showed a slow response time
(as long as 1 ps) with an intermediate structure displaying peaks
at 13.75$^{\circ}$ and 13.8$^{\circ}$ \cite{Cavalleri-1}. In this
study the metal phase peak at 13.8$^{\circ}$ appeared after times
longer than 300 fs after a 50 fs x-ray irradiation. This indicated
that the SPT is continuous. As for the MIT, the rate of change in
the reflectivity due to the increased free carriers was observed
to be as short as 80 fs at the most \cite{Cavalleri-2,Rini}. It
was suggested that this shorter response time might be due to a
structural bottleneck since the authors of the works had assumed
the concurrence of the MIT with the SPT
\cite{Cavalleri-1,Cavalleri-2}. The response time difference
between 300 fs and 80 fs supports the idea that the MIT and the
SPT cannot occur simultaneously. This is consistent with the
results shown in the present work.

We will now explain the origin of the MCM phases in Fig. 1. The
critical hole density required to induce the MIT in VO$_2$ was
theoretically predicted by the hole-driven MIT theory
\cite{Kim-1,Kim-4}, and has been known as n$_c\approx$0.018$\%$
\cite{Kim-2}. The critical hole density is given by
n$_c$(T,Photo)=n(T)+n(Photo) where n$_c$(T,Photo) is the hole
density excited by both light and temperature, and n(T) and
n(Photo) are the hole carrier densities excited by temperature and
light, respectively. External excitations by temperature,
pressure, chemical doping and light generate the transition to the
MCM phase according to the MIT condition. If there is no external
excitation, only n(T) can induce the MCM phase.

This scenario is based on the hole-driven MIT theory, which
explains the breakdown of the critical on-site Coulomb energy by
hole doping doping of a low concentration into the valence band of
the Mott insulator \cite{Kim-1,Kim-4}. It predicts that the MIT
can be switched on or off by the doping or de-doping of the
valence band with a low concentration of holes \cite{Kim-4}.
Recently there have been several experimental reports confirming
that the abrupt MIT is induced by holes
\cite{Kim-2,Cavalleri-2,Basov,Lee}. We have experimentally
demonstrated that the first-order MIT occurs with the doping of
the valence band to a critical hole density \cite{Kim-2}. The MIT
induced by temperature is identical to the abrupt first-order MIT
observed by an external electric field \cite{Kim-2}, since MITs
can be driven only by holes, irrespective of the method of
excitation. Moreover, the MIT differs from the transition driven
by the SPT, which has lead us to believe that VO$_2$ is a Peierls
insulator \cite{Goodenough,Wentzcovitch,Haverkort,Cavalleri-2}.
The first-order MIT is quite different from the Mott-Hubbard
continuous MIT in which the density of states on the Fermi surface
gradually decreases as the on-site Coulomb potential increases
\cite{Zhang}.

The MCM phase is supposed to have a maximum conductivity, related
to the maximum effective mass, $m^{\ast}$, near the MIT as
${\sigma\propto}(m^{\ast}/m)^2$ \cite{Kim-2,Kim-1,Kim-4}. The
maximum effective mass can be regarded as a diverging true
effective mass in the Brinkman-Rice picture \cite{Brinkman}. It
was suggested that the metal phase is correlated by I-V
measurement \cite{Kim-4} and optical measurements
\cite{Qazilbash}. The monoclinic T phase instead of M$_2$ in
VO$_2$ \cite{Pouget,Rice,Tomakuni} can be classified as a
correlated paramagnetic Mott insulator with the equally spaced V-V
chain, on the basis of the jump between the T and MCM phases. The
T phase is defined as a semiconductor or insulator phase before
the transition from insulator to metal occurs near 340 K.

In conclusion, the first-order MIT is driven not by the SPT but by
hole carriers in VO$_2$ and occurs between T and MCM. The
monoclinic and correlated metal phase can be regarded as a
non-equilibrium state because the MCM phase exists at the
divergence in the hole-driven MIT theory and the Brinkman-Rice
picture. The characteristics of the MCM phase need to be studied
in more depth.

%\section*{ACKNOWLEDGEMENTS}
We acknowledge Prof. D. N. Basov and Dr. M. M. Qazilbash for
valuable comments. This research was performed by a project of
High Risk High Return in ETRI.


\begin{references}
\bibitem{Laser} We observed a laser spot formed by a high laser power on a VO$_2$ film below 68$^{\circ}$C in a micro-Raman experiment.
The laser spot was not observed in a low laser power and is a
cause of local heating decreasing the observed $T_{SPT}$ such as
the case of 100 mW in Fig. 1 (c).
\end{references}
\end{document}